\documentclass[12pt,preprint]{aastex}

\newcommand{\kpc}{\,{\rm kpc}}

\received{}
\revised{}
\accepted{}

\shorttitle{Galaxy Shape Dependence}
\shortauthors{Vincent \& Ryden}

\begin{document}
\title{The Dependence of Galaxy Shape on Luminosity
and Surface Brightness Profile}
\author{R. Anthony Vincent}
\affil{Department of Physics, The Ohio State University,
Columbus, OH 43210}
\email{vincent.59@osu.edu}
\and
\author{Barbara S. Ryden }
\affil{Department of Astronomy, The Ohio State University,
Columbus, OH 43210}
\email{ryden@astronomy.ohio-state.edu}

\begin{abstract}

For a sample of 96{,}951 galaxies from the Sloan Digital Sky Survey
Data Release 3, we study the distribution of apparent axis ratios
as a function of $r$-band absolute magnitude and surface brightness
profile type. We use the parameter \texttt{fracDeV} to quantify the
profile type ($\texttt{fracDeV} = 1$ for a pure de Vaucouleurs profile;
$\texttt{fracDeV} = 0$ for a pure exponential profile). When the
apparent axis ratio $q_{\rm am}$ is estimated from the moments of
the light distribution, the roundest galaxies are very bright
($M_r \sim -23$) de Vaucouleurs galaxies and the flattest are
modestly bright ($M_r \sim -18$) exponential galaxies. When the
axis ratio $q_{25}$ is estimated from the axis ratio of the 25
mag/arcsec$^2$ isophote, we find that de Vaucouleurs galaxies,
at this low surface brightness, are flatter than exponential
galaxies of the same absolute magnitude. For a given surface
brightness profile type, very bright galaxies are rounder, on
average, than fainter galaxies. We deconvolve the distributions
of apparent axis ratios to find the distribution of the
intrinsic short-to-long axis ratio $\gamma$, making the
assumption of constant triaxiality $T$. For all profile
types and luminosities, the distribution of axis ratios
is inconsistent with a population of oblate spheroids, but
is usually consistent with a population of prolate spheroids.
Bright galaxies with a de Vaucouleurs
profile ($M_r \leq -21.84$, $\texttt{fracDeV} > 0.9$)
have a distribution of $q_{\rm am}$ that is consistent
with triaxiality in the range $0.4 \la T \la 0.8$, with
mean axis ratio $0.66 \la \langle \gamma \rangle \la 0.69$. The
fainter de Vaucouleurs galaxies are best fit with
prolate spheroids ($T = 1$) with mean axis ratio
$\langle \gamma \rangle \approx 0.51$.

\end{abstract}


\section{INTRODUCTION}
\label{sec-intro}

The galaxy classification scheme of \citet{hu26} has proved durably useful.
As modified and extended by \citet{de59}, it is still the standard method
for classifying low-redshift galaxies with high surface brightness.
The Hubble classification scheme was originally based on the
appearance of galaxies on photographic plates. Elliptical galaxies
have smooth elliptical isophotes; spiral galaxies have spiral arms
that wind outward from a central bulge or bar. It was later
discovered that for luminous galaxies, the surface brightness
profile is strongly correlated with the Hubble type. If the
surface brightness $I$ is measured along the major axis of
a galaxy's image, it is found that bright elliptical galaxies
have surface brightness profiles that are well fit by a
de Vaucouleurs, or $R^{1/4}$ law, for which $\log I \propto
- R^{1/4}$ \citep{de48}. By contrast, the azimuthally averaged
surface brightness profile of a spiral galaxy, outside
its central bulge, is typically well fit by
an exponential law, $\log I \propto - R$ \citep{fr70}.
It was also eventually realized that galaxies of different
Hubble type have different kinematic properties. The disks
of spiral galaxies are rotationally flattened, with stars and
gas on nearly circular orbits with little random motion. Bright
elliptical galaxies ($M_B \la -20$), by contrast, are slowly
rotating and are supported mainly by their anisotropic velocity
dispersion.

One shortcoming of the Hubble classification scheme, imposed
by necessity, is that elliptical galaxies are classified by
their apparent two-dimensional shape, seen in projection on
the sky, rather than their intrinsic three-dimensional shape.
Consider an idealized galaxy whose surfaces of constant luminosity
density are concentric, coaxial, similar ellipsoids, with principal
axes of lengths $a \geq b \geq c$; the shape of the galaxy can then
be described by the two axis ratios $\beta \equiv b/a$ and
$\gamma \equiv c/a$. Equivalently, the shape can be described by
the two numbers $\gamma$ and $T$, where the triaxiality parameter
$T$ is given by the relation $T \equiv (1-\beta^2) / (1-\gamma^2)$.
If the ellipsoidal galaxy is seen in projection, though, its isophotes
will be concentric, coaxial, similar ellipses. The shape of the
projected image can then be described by the single axis ratio
$q \equiv B/A$, where $A$ and $B$ are the major and minor axis
length of any isophote.

Although knowing the apparent axis ratio $q$ is not, by itself,
sufficient to determine the intrinsic axis ratios $\beta$ and $\gamma$,
the three-dimensional shape of galaxies is not beyond all conjecture.
Two approaches to determining the three-dimensional shape of galaxies
have been used. First, the intrinsic shape of an individual galaxy
can be modeled if kinematic data are available in addition to photometric
data \citep{bi85,fi91}. However, accurate galaxy modeling requires either
multiple long-slit position angles \citep{st94} or integral-field
spectroscopy \citep{se04}, and is best applied to systems with high
surface brightness and large angular size. The second approach, which
can be used in the absence of kinematic information, is to make statistical
estimates of the distribution of $\beta$ and $\gamma$ for a large sample
of galaxies. Early estimates of the intrinsic shape distribution made
the assumption that elliptical galaxies were oblate spheroids, with
$\beta = 1$ \citep{hu26,sf70}. More recent studies, using accurate
surface photometry, reveal that the distribution of apparent shapes
for ellipticals cannot be explained by a population of randomly oriented
oblate spheroids; it can easily be explained, however, by a population
of triaxial ellipsoids \citep{fv91, lm92, ry92, tm95, ar02}.

In this paper, we use the photometry-only approach to constraining
the intrinsic shapes of galaxies, using the Sloan Digital Sky
Survey Data Release 3 (SDSS DR3) as our source of galaxy photometry.
Previous studies using the SDSS Early Data Release and Data Release 1
studied the axis ratios of galaxies best fit by de Vaucouleurs profiles
\citep{ar02} and those best fit by exponential profiles \citep{ry04}.
In this paper, we more closely examine the relation between surface
brightness profile and intrinsic shape. In addition, we determine, for
each profile type, the dependence of intrinsic shape on galaxy luminosity.
For elliptical galaxies, the distribution of apparent shapes is known
to depend on absolute magnitude. Elliptical galaxies with $M_B \la -20$
are rounder on average than fainter ellipticals \citep{tm96}; for
a typical elliptical galaxy color of $\bv \approx 0.9$ \citep{rh94},
the dividing luminosity corresponds to $M_V \approx -20.9$. In this
paper, given the large sample size provided by the SDSS DR3, we can
look for a similar shape dichotomy not only among galaxies with
de Vaucouleurs profiles, but among galaxies with a variety of
surface brightness profile types.

In \S\ref{sec-data} of this paper, we describe the SDSS and the
methods by which we determine the apparent axis ratio of the galaxies
in our sample. In \S\ref{sec-app}, we examine how the apparent
axis ratios depend on the galaxy profile type and the galaxy luminosity,
then use nonparametric kernel estimators to determine the
distribution of apparent axis ratios for different samples
of galaxies, subdivided by luminosity and profile type.
In \S\ref{sec-int}, we invert the distribution of apparent axis
ratios to find the distribution of intrinsic axis ratios, assuming
galaxies all have the same trixiality parameter $T$. In addition to
looking at purely oblate galaxies ($T = 0$) and purely prolate
galaxies ($T = 1$), we also examine the results for triaxial
galaxies with $T = 0.2$, $T = 0.4$, $T = 0.6$, and $T = 0.8$.
in \S\ref{sec-dis}, we discuss the implications of the observed
galaxy shapes for different scenarios of galaxy formation and
evolution.

\section{DATA}
\label{sec-data}

The Sloan Digital Sky Survey \citep{yo00, st02} will, when complete,
provide a map of nearly one-fourth of the celestial sphere. A
CCD mosaic camera \citep{gu98} images the sky in five
photometric bands ($ugriz$; \citet{fu96, sm02}). The Main
Galaxy Sample (MGS) of the SDSS \citep{st02} will eventually contain
$\sim 10^6$ galaxies with $r \leq 17.77$; the mean redshift of
galaxies in the MGS is $\sim 0.1$, determined from a follow-up
spectroscopic survey. The SDSS Data Release 3, issued to the
astronomical community in 2004 October, contains 5282 square
degrees of imaging data and 4188 square degrees of spectroscopic
data (\citet{ab05}; see also \citet{st02}, \citet{ab03}, and \citet{ab04}).

The SDSS DR3 data processing pipeline provides a morphological
star/galaxy separation, with extended objects being classified
as `galaxies' and point-like objects being classified as `stars'.
For each galaxy, in each photometric band, a pair of models are
fitted to the two-dimensional galaxy image. The first model
has a de Vaucouleurs surface profile \citep{de48}:
\begin{equation}
I(R) = I_e \exp \left( -7.67 [ (R/R_e)^{1/4} - 1 ] \right) \ ,
\end{equation}
truncated beyond $7 R_e$ to go smoothly to zero at
$8 R_e$. The second model has an exponential profile:
\begin{equation}
I(R) = I_e \exp \left( -1.68 [ R/R_e - 1 ] \right) \ ,
\end{equation}
truncated beyond $3 R_e$ to go smoothly to zero
at $4 R_e$. For each model, the apparent axis ratio $q_m$
and the phase angle $\varphi_m$ are assumed to be constant
with radius. The parameters $q_m$, $\varphi_m$, $R_e$, and
$I_e$ are varied to give the best $\chi^2$ fit to the
galaxy image, after convolution with a double-Gaussian
fit to the point spread function.

The SDSS DR3 pipeline also takes the best-fit
exponential model and the best-fit de Vaucouleurs model
and finds the linear combination of the two that best fits
the galaxy image. The fraction of the total flux contributed
by the de Vaucouleurs component is the parameter \texttt{fracDeV},
which is constrained to lie in the interval $0 \leq
\texttt{fracDeV} \leq 1$. The \texttt{fracDeV} parameter delineates
a continuum of surface brightness profile types, from the
pure de Vaucouleurs ($\texttt{fracDeV} = 1$) to the pure exponential
($\texttt{fracDeV} = 0$).

The SDSS DR3 databases provide many different measures of the
apparent axis ratio $q$ of each galaxy in each of the five
photometric bands. In this paper, we will use the $r$ band data,
at an effective wavelength of 6165{\AA}. (We also repeated our analysis
at $g$ and $i$, the other two bands with high signal-to-noise, and
did not find significantly different results from our $r$ band analysis.)
Two measures of the apparent axis ratio are given by the best-fitting
axis ratios $q_m$ for the de Vaucouleurs and exponential models.
However, the algorithm for fitting the models introduces quantization
in the distribution of $q_m$. Because of this artificial quantization,
we do not use the model axis ratios as estimates of the true apparent
shapes of galaxies.

A useful measure of the apparent shape in the outer regions of galaxies
is the axis ratio of the 25 mag arcsec$^{-2}$ isophote. The SDSS DR3
data pipeline finds the best fitting ellipse to the 25 mag arcsec$^{-2}$
isophote in each band; the semimajor axis and semiminor axis of this
isophotal ellipse are $A_{25}$ and $B_{25}$. The isophotal axis ratio
$q_{25} \equiv B_{25}/A_{25}$ then provides a measure of the apparent
galaxy shape at a few times the effective radius. For galaxies in
our sample with $\texttt{fracDeV} = 1$, the mean and standard deviation
of $A_{25}/R_e$ are $3.12 \pm 1.00$; for galaxies with $\texttt{fracDeV}
= 0$, the mean and standard deviation are $A_{25}/R_e = 2.40 \pm 0.36$.

Another measure of the apparent shape is $q_{\rm am}$, the axis ratio
determined by the use of adaptive moments of the galaxy's light. The
method of adaptive moments determines the $n$th order moments of a
galaxy image, using an elliptical weight function whose shape matches
that of the image \citep{bj02,hs03}. The SDSS DR3 adaptive moments use
a weight function $w (x,y)$ that is a Gaussian matched to the size and
ellipticity of the galaxy image $I(x,y)$. The adaptive first order moments,
\begin{equation}
x_0 = {\int x w(x,y) I(x,y) dx dy \over \int w(x,y) I(x,y) dx dy}
\end{equation}
and
\begin{equation}
y_0 = {\int y w(x,y) I(x,y) dx dy \over \int w(x,y) I(x,y) dx dy} \ ,
\end{equation}
tell us the `center of light' of the galaxy's image. With this knowledge,
we can compute the adaptive second order moments:
\begin{equation}
M_{xx} = {\int (x-x_0)^2 w(x,y) I(x,y) dx dy \over \int w(x,y) I(x,y)
dx dy} \ ,
\end{equation}
and so forth. The SDSS DR3 provides for each image the values of the
parameters $\tau = M_{xx} + M_{yy}$, $e_+ = (M_{xx}-M_{yy})/\tau$,
and $e_\times = 2 M_{xy} / \tau$. The adaptive second moments can
be converted into an axis ratio using the relation
\begin{equation}
q_{\rm am} = \left( {1-e \over 1+e} \right)^{1/2} \ ,
\end{equation}
where $e = (e_+^2 + e_\times^2)^{1/2}$.

The adaptive moments axis ratio $q_{\rm am}$, computed in the
manner given above, is not corrected for the effects of seeing.
The SDSS DR3 also provides the fourth order adaptive moments of the
galaxy image, and the adaptive moments $\tau_{\rm psf}$, $e_{+,{\rm psf}}$,
and $e_{\times,{\rm psf}}$ of the point spread function at the galaxy's
location. These moments can be used to correct for the smearing and
shearing due to seeing; such corrections are essential for studying
the small shape changes resulting from weak lensing \citep{bj02,hs03}.
However, in this paper we will only look at well-resolved galaxies
for which the seeing corrections are negligibly small.

Our full sample of galaxies consists of those objects in the SDSS DR3
spectroscopic sample which are flagged as galaxies and which have
spectroscopic redshifts $z > 0.004$ (to eliminate contaminating foreground
objects) and $z < 0.12$ (to reduce the possibility of weak lensing
significantly distorting the observed shape). To ensure that galaxies
were well resolved, we also required $\tau > 7 \tau_{\rm psf}$. The
absolute magnitude $M_r$ of each galaxy was computed from the Petrosian
$r$ magnitude and the spectroscopic redshift, assuming a Hubble constant
$H_0 = 70 {\rm\,km}{\rm\,s}^{-1}{\rm\,Mpc}^{-1}$ in a flat universe with
$\Omega_{\rm m,0} = 0.3$ and $\Omega_{\Lambda,0} = 0.7$. No K-correction
was applied to the data; for galaxies with normal colors at low redshift
the K-correction in the $r$ band is small -- less than 0.2 mag for an
average elliptical galaxy at $z = 0.12$, and even less for galaxies
at smaller redshifts or with bluer colors \citep{fu95}.

The total number of galaxies in our full sample,
selected in this way, is $N_{\rm gal} = 96{,}951$. Of the
full sample, only 919 galaxies have $\texttt{fracDeV} = 1$,
and 26{,}994 have $\texttt{fracDeV} = 0$. The remainder,
constituting 71\% of the full sample,
are best fit by a combination of a de Vaucouleurs and
exponential model, with $0 < \texttt{fracDeV} < 1$.

\section{APPARENT SHAPES}
\label{sec-app}

It remains to be demonstrated that \texttt{fracDeV} is
a useful parameter for describing the surface brightness
profile type of galaxies. After all, a linear combination
of the best de Vaucouleurs fit and the best exponential
fit will not, in general, be the best possible de Vaucouleurs plus
exponential fit.  An alternative method of describing surface
brightness profiles is by fitting a S\'ersic profile \citep{se68}:
\begin{equation}
\log I \propto - (R / R_e )^{1/n} \ ,
\end{equation}
where $n$ can be an arbitrary number. This fitting method was used,
for instance, by \citet{bl03} in their study of SDSS galaxies. If,
in fact, galaxies are well described by S\'ersic profiles, then
\texttt{fracDeV} is a useful surrogate for the S\'ersic index $n$.
Consider, for instance, a galaxy whose surface brightness is
perfectly described by a S\'ersic profile of index $n = 2$ and
effective radius $R_e = X$. The best-fitting exponential model
for this galaxy (fitting in the radial region $0.1 X \leq R \leq 10 X$)
has $R_e = 1.10 X$; the best-fitting de Vaucouleurs model has
$R_e = 0.80 X$. Combining the models gives $\texttt{fracDeV} = 0.51$
for this $n = 2$ S\'ersic galaxy. A similar fit to an $n = 3$
S\'ersic galaxy yields $\texttt{fracDeV} = 0.83$. In general,
if galaxies have S\'ersic profiles with $1 \leq n \leq 4$, then
\texttt{fracDeV}, as computed by the SDSS DR3 pipeline, is a
monotonically increasing function of the S\'ersic index $n$,
and thus can be used as a surrogate for $n$. 

\begin{figure}
\plotone{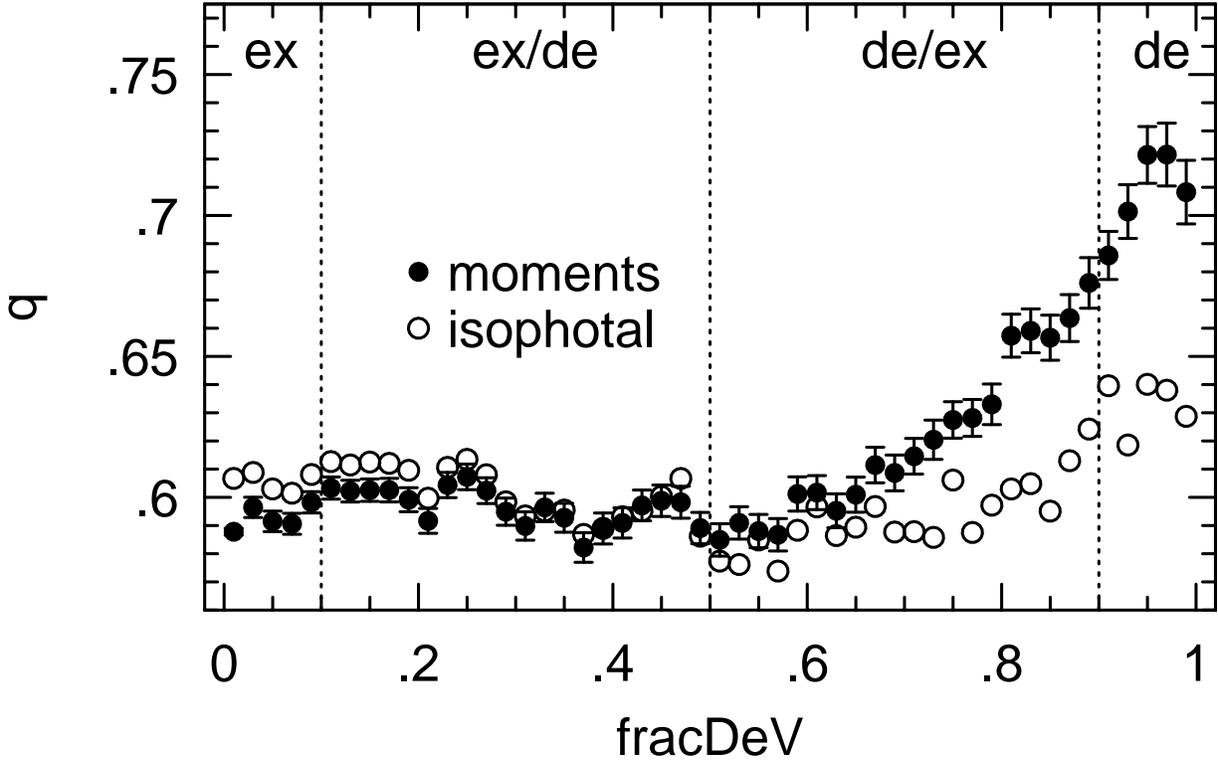}
\caption{Mean apparent axis ratio $\langle q \rangle$ as a function of
the SDSS fitting parameter \texttt{fracDeV}. Filled
circles indicate the axis ratio estimated using the
adaptive moments technique; open circles indicate the
axis ratio of the 25 mag/arcsec$^2$ isophote. The error
bars on the filled circles indicate the estimated error of
the mean; the error bars on the open circles (not shown for
clarity) are of similar size.
}
\label{fig:fdev}
\end{figure}

A plot of the mean axis ratio $\langle q \rangle$ as a function
of the parameter \texttt{fracDeV} is shown in Figure~\ref{fig:fdev}.
The mean adaptive moments axis ratio $\langle q_{\rm am} \rangle$,
indicated by the filled circles, shows an increasing trend with
\texttt{fracDeV}, from $\langle q_{\rm am} \rangle = 0.59$ for
galaxies with purely exponential profiles ($\texttt{fracDeV} = 0$)
to $\langle q_{\rm am} \rangle = 0.74$ for galaxies with pure
de Vaucouleurs profiles ($\texttt{fracDeV} = 1$). However, the
trend in $\langle q \rangle$ is not a linear one; for galaxies
with $\texttt{fracDeV} \la 0.5$, the value of $\langle q \rangle$
is nearly constant at $\langle q \rangle \approx 0.6$; it is only
at $\texttt{fracDeV} \ga 0.5$ that $\langle q \rangle$ shows an
increasing trend with \texttt{fracDeV}. The mean isophotal
axis ratio $\langle q_{25} \rangle$, indicated by the open
circles in Figure~\ref{fig:fdev}, shows less of a trend with
\texttt{fracDeV}. Except in the case of nearly pure de Vaucouleurs
profiles ($\texttt{fracDeV} \ga 0.9$), the axis ratio of a galaxy
in its outer regions doesn't seem to depend on its surface brightness
profile.

For convenience in analysis, we have divided our galaxy sample into
four classes, based on the value of \texttt{fracDeV}. Galaxies with
$\texttt{fracDeV} \leq 0.1$, corresponding to a S\'ersic index
$n \la 1.2$, are called `ex' galaxies; there are $N_{\rm ex} =
44{,}289$ `ex' galaxies in our sample. Galaxies with $0.1 <
\texttt{fracDeV} \leq 0.5$, corresponding to $1.2 \la n \la 2.0$,
are labeled `ex/de' galaxies ($N_{\rm ex/de} = 36{,}645$).
Galaxies with $0.5 < \texttt{fracDeV} \leq 0.9$, corresponding
to $2.0 \la n \la 3.3$, are labeled `de/ex' galaxies
($N_{\rm de/ex} = 13{,}780$). Finally, galaxies with $\texttt{fracDeV}
> 0.9$, corresponding to $n \ga 3.3$, are called `de' galaxies.
The small fraction of `de' galaxies in our sample ($N_{\rm de} =
2237$) is partly due to the fact that the centrally concentrated
de Vaucouleurs galaxies are less likely to satisfy our resolution
criterion, and partly due to the fact that galaxies with high
S\'ersic indices are intrinsically rare; \citet{bl03}
estimated that only $\sim 5\%$ of the SDSS galaxies in their
sample had S\'ersic index $n > 3$.

The `de' galaxies are rounder in their central regions than
in their outer regions: $\langle q_{\rm am} - q_{25} \rangle =
0.083$. This is consistent with the `de' galaxies being relatively
bright elliptical galaxies, for which the isophotal axis ratios tend
to decrease with increasing semimajor axis length \citep{rf01}. By
contrast, the `ex' galaxies are actually slightly flatter in their
central regions than in their outer regions: $\langle q_{\rm am} -
q_{25} \rangle = -0.017$.

\begin{figure}
\plotone{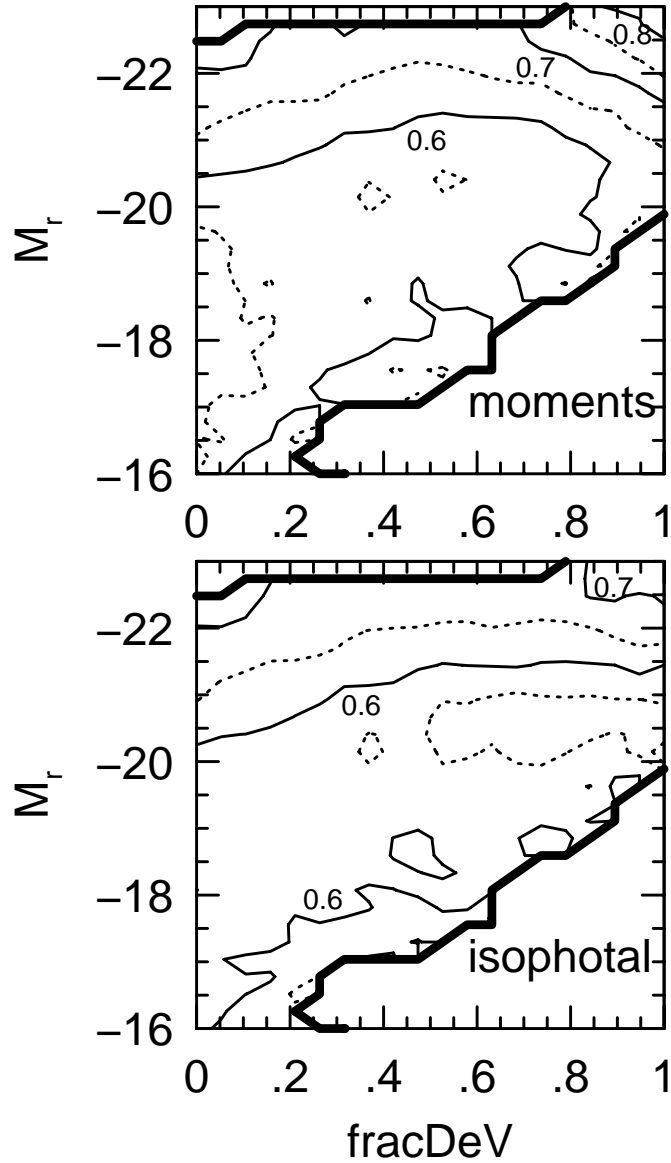}
\caption{\emph{Upper panel:} Mean apparent axis ratio $q_{\rm am}$,
estimated using adaptive moments, as a function of
absolute magnitude $M_r$ and fitting parameter \texttt{fracDeV}.
Averages were computed in bins of width 0.25 mag in $M_r$ and
0.05 in \texttt{fracDeV}; the heavy line excludes the region with
fewer than 25 galaxies per bin. Contours are drawn at $q = 0.6$, $0.7$,
and $0.8$ (solid lines) and at $q = 0.55$, $0.65$, and $0.75$
(dotted lines). \emph{Lower panel:} Same as the upper panel,
using the isophotal axis ratio $q_{25}$ as the shape estimate.
}
\label{fig:fdevmag}
\end{figure}

For a given surface brightness profile type, there exists a relation
between absolute magnitude and apparent axis ratio. Contour plots
of mean apparent axis ratio as a function of both \texttt{fracDeV}
and $M_r$ are given in Figure~\ref{fig:fdevmag}. (To give a feel for
the absolute magnitude scale, fitting a Schechter
function to the luminosity function of SDSS
galaxies yields $M_{*,r} \approx -21.4$ \citep{na03}.) The upper panel
of the figure shows that the trend in $\langle q_{\rm am} \rangle$
runs from the flattest galaxies at $\texttt{fracDeV} \approx 0$
and $M_r \approx -18$, where $\langle q_{\rm am} \rangle \approx 0.52$,
to the roundest galaxies at $\texttt{fracDeV} \approx 1$ and
$M_r \approx -23$, where $\langle q_{\rm am} \rangle \approx 0.83$.
This result is not surprising, since moderately bright galaxies
with exponential profiles are intrinsically flattened disk galaxies,
while extremely bright galaxies with de Vaucouleurs profiles are
intrinsically nearly spherical giant elliptical galaxies \citep{tm96}.
More surprising are the results shown in the bottom panel of
Figure~\ref{fig:fdevmag}, which shows $\langle q_{25} \rangle$, the
mean isophotal axis ratio. Here, we find that the apparently
flattest galaxies, as measured by $q_{25}$, are not exponential
galaxies, but galaxies with $\texttt{fracDeV} \approx 0.7$ (corresponding
to S\'ersic index $n \approx 2.5$) and $M_r \approx -20.5$; these
galaxies have $\langle q_{25} \rangle \approx 0.53$. The roundest
galaxies, measured by $\langle q_{25} \rangle$, are not bright
de Vaucouleurs galaxies, but bright exponential galaxies; the
maximum value of $\langle q_{25} \rangle$ is $\approx 0.74$, at
$\texttt{fracDeV} \approx 0$, $M_r \approx -22.5$. Note also that
for bright galaxies ($M_r \la -21$), the contours of constant
$\langle q_{25} \rangle$ in Figure~\ref{fig:fdevmag}
are nearly horizontal. That is, among
bright galaxies, the flattening of the outer isophotes doesn't
depend strongly on the surface brightness profile type.

\begin{figure}
\plotone{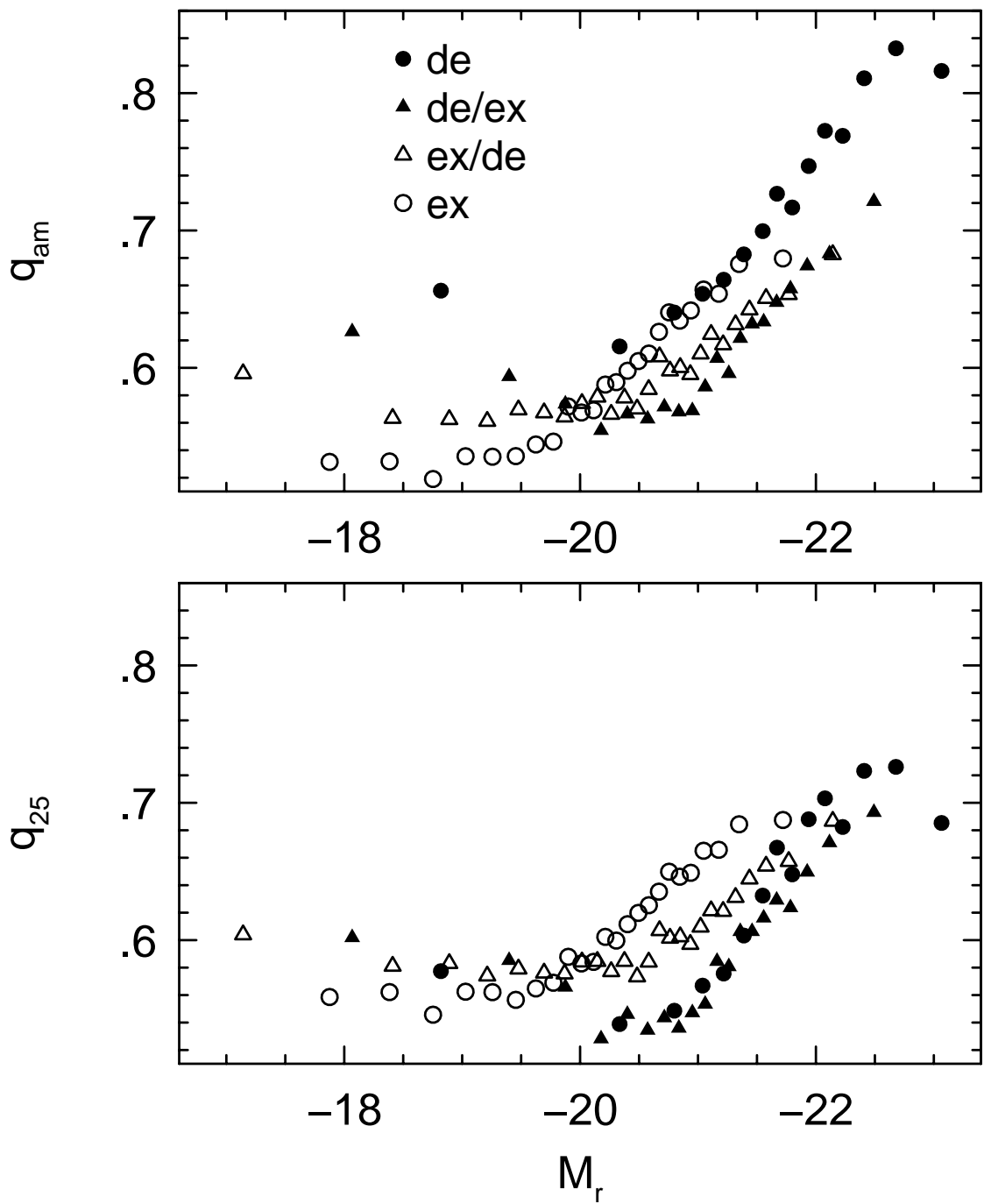}
\caption{\emph{Upper panel:} Mean apparent axis ratio $q_{\rm am}$,
estimated using the adaptive moments technique, as a function of
absolute magnitude. The four different galaxy profile types (de, de/ex,
ex/de, ex) are described in the text.
\emph{Lower panel:} Same as the upper panel, using the
apparent axis ratio $q_{25}$ of the 25 mag/arcsec$^2$ isophote.
}
\label{fig:mag}
\end{figure}

A view of the dependence of $\langle q \rangle$ on absolute
magnitude for each of our four profile types, `ex', `ex/de',
'de/ex', and `de', is given in Figure~\ref{fig:mag}. In the
upper panel, which shows $\langle q_{\rm am} \rangle$, note
that for each profile type, there is a critical absolute magnitude
$M_{\rm crit}$ at which $\langle q_{\rm am} \rangle$ is at a minimum.
This critical absolute magnitude ranges from $M_{\rm crit} \sim -20.6$
for the `de' galaxies to $M_{\rm crit} \sim -19.4$ for the `ex' galaxies.
At $M_r < M_{\rm crit}$, the value of $\langle q_{\rm am} \rangle$ increases
relatively rapidly with increasing luminosity; at $M_r > M_{\rm crit}$, the
value of $\langle q_{\rm am} \rangle$ increases less rapidly
with decreasing luminosity. At a fixed absolute magnitude, the
average axis ratio of `de' galaxies is always greater than
that of `ex' galaxies; however, for $M_r \la -20$, the flattest
galaxies, on average, at a given absolute magnitude are not the
`ex' galaxies, but those with the mixed `de/ex' and `ex/de'
profile types. The bottom panel of Figure~\ref{fig:mag} shows
$\langle q_{25} \rangle$ versus $M_r$ for the different
profile types. In the interval $-20 \la M_r \la -22$, we see
that `ex' galaxies have \emph{larger} values of $\langle
q_{25} \rangle$ than galaxies of other profile types.

\citet{tm96} divided elliptical galaxies into two classes;
galaxies brighter than $M_B \approx -20$ are rounder on average
than fainter ellipticals. Using a typical color for elliptical
galaxies of $\bv \approx 0.9$, this corresponds to an absolute
magnitude in the $r$ band of $M_r \approx -21.2$, using the
transformation $r = B - 1.44 (B-V) + 0.12$ \citep{sm02}.
In our data, we do not see an abrupt jump in $\langle q_{\rm am}
\rangle$ for `de' galaxies at $M_r = -21.2$, but rather a more gradual
increase between $M_r \approx -20.6$ and $M_r \approx -22.7$.
In addition, it is true for any profile type that highly luminous galaxies
are rounder, on average, than less luminous galaxies.

Knowing the mean shape as a function of $M_r$ for each profile type
(`de', `de/ex', `ex/de', 'ex') does not tell us the complete distribution
of axis ratios, $f(q)$. In order to compare the distribution of axis
ratios for highly luminous galaxies with that for less luminous galaxies,
we determined, for each profile type, the dividing absolute magnitude
$M_0$ such that the distribution $f(q_{\rm am})$ for galaxies with
$M_r \leq M_0$ is maximally different from the distribution $f(q_{\rm am})$
for galaxies with $M_r > M_0$. We measure the difference between the
distributions by applying a Kolmogorov-Smirnov (K-S) test, and finding the
value of $M_0$ that minimizes the K-S probability
$P_{\rm KS}$. The values of the dividing magnitude $M_0$ for each profile
type, and the associated probability $P_{\rm KS}$, are given in
Table~\ref{tab:bridim}. (Using a Student's t-test for the significance
of the difference of the means yields similar values of $M_0$,
differing by less than 0.2 mag from the values found using the
K-S test.)

To estimate the distribution of apparent axis ratios $f(q)$ for each of
our galaxy subsamples, we use a nonparametric kernel technique
\citep{vi94,tm95,ry96}. Given a sample of $N$ axis ratios, $q_1$,
$q_2$, $\dots$, $q_N$, the kernel estimate of the frequency distribution
$f(q)$ is
\begin{equation}
\hat{f} (q) = {1 \over N} \sum_{i=1}^N {1 \over h_i}
K \left( {q - q_i \over h_i} \right) \ ,
\label{eq:kernel}
\end{equation}
where $K(x)$ is the kernel function, normalized so that
\begin{equation}
\int_{-\infty}^\infty K(x) dx = 1 \ .
\end{equation}
To ensure that our estimate $\hat{f}$ is smooth, we use
a Gaussian kernel, with $K(x) \propto \exp (-x^2/2)$. The parameter $h_i$ in
equation~(\ref{eq:kernel}) is the kernel width; too small a width introduces noise
into the estimate, while too great a kernel width produces excessive smoothing.
We choose the kernel width $h_i$ by using the standard adaptive two-stage
estimator of \citet{ab82}. In this technique, an initial estimate $\hat{f}_0$
is made using a global fixed kernel width 
\begin{equation}
h = 0.9 A / N^{0.2} \ ,
\label{eq:hwidth}
\end{equation}
with $A = \min (\sigma, Q_4 / 1.34 )$, where $\sigma$ is the
standard deviation and $Q_4$ is the interquartile range
of the axis ratios. For samples that are not strongly skewed,
this formula minimizes the mean square error \citep{si86,vi94}.
The final estimate $\hat{f}$ is then computed, using at each data point $q_i$
the kernel width
\begin{equation}
h_i = h \left[ G / \hat{f}_0 (q_i) \right]^{1/2} \ ,
\end{equation}
where $G$ is the geometric mean of $\hat{f}_0$ over all values of $i$.

The axis ratio for a galaxy must lie in the range $0 \leq q \leq 1$. To
ensure that $\hat{f} = 0$ for $q < 0$ and $q > 1$, we impose reflective
boundary conditions at $q = 0$ and $q = 1$. In practice, this means replacing
the simple Gaussian kernel with the kernel
\begin{equation}
K_{\rm ref} = K \left( {q - q_i \over h_i} \right) + K \left( {- q - q_i \over
h_i} \right) + K \left( {2 - q - q_i \over h_i} \right) \ .
\end{equation}
Although use of reflective boundary conditions ensures that $\hat{f}$ remains
within bounds, it imposes the possibly unphysical constraint that $d \hat{f}
/ dq = 0$ at $q = 0$ and $q = 1$.

\begin{figure}
\plotone{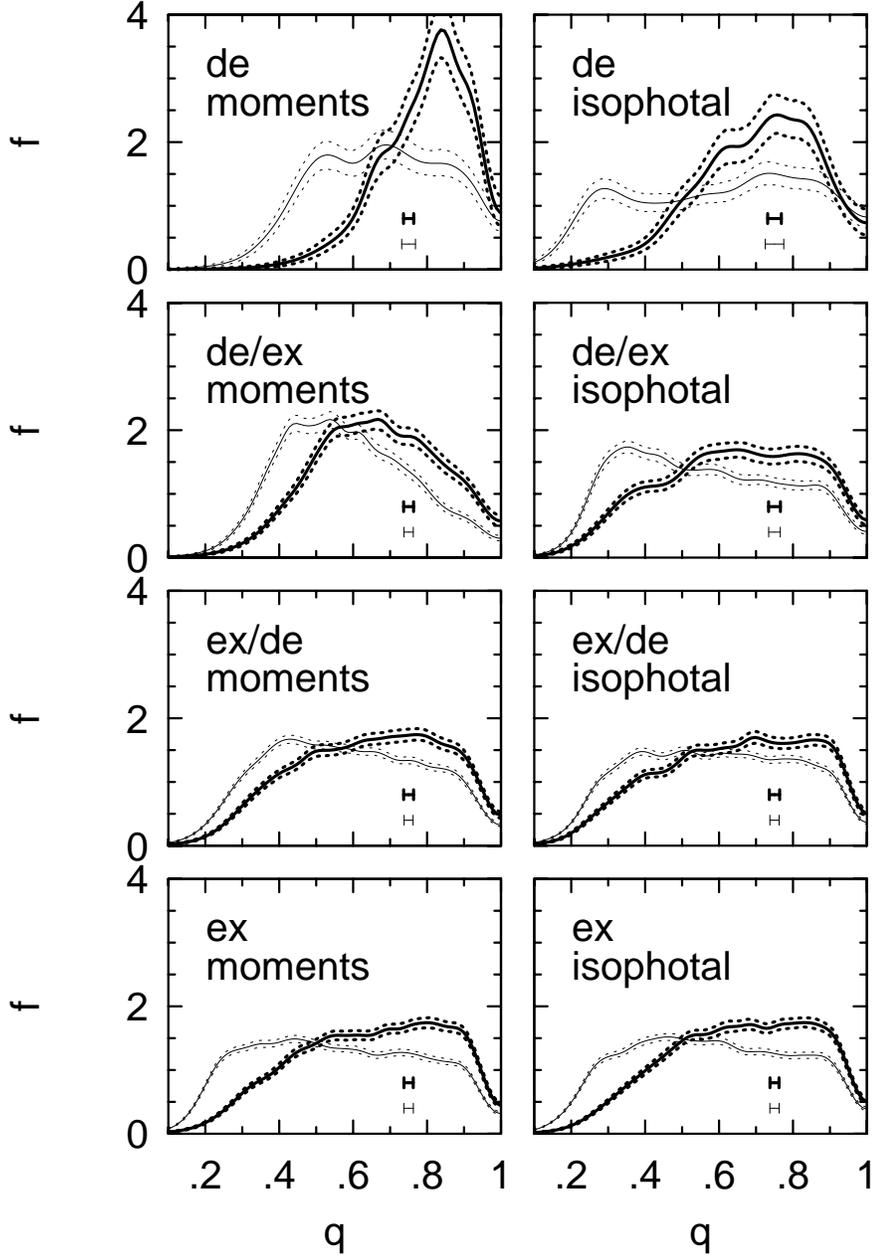}
\caption{
\emph{Left column:} Distribution of the adaptive moments
axis ratio $q_{\rm am}$. From top to bottom, the results are
shown for profile types `de', `de/ex', `ex/de', and `ex'.
In each panel, the heavy line is the distribution for bright
galaxies, and the light line is the distribution for faint galaxies.
The dotted lines indicate the 98\% confidence interval found by
bootstrap resampling. The horizontal error bars indicate the
kernel width $h$.
\emph{Right column:} The same as the left column, but using the
25 mag/arcsec$^2$ isophotal axis ratio, $q_{25}$, as the shape estimator.
}
\label{fig:apparent}
\end{figure}

The estimated distributions $\hat{f} (q_{\rm am})$ are shown in the
left column of Figure~\ref{fig:apparent}, for the `de', `de/ex', `ex/de',
and `ex' galaxies; the equivalent distributions $\hat{f} (q_{25})$
for the isophotal shapes are shown in the right column. In each panel
of Figure~\ref{fig:apparent}, the heavier line indicates the
shape distribution for the bright galaxies, and the lighter line
indicates the distribution for the fainter galaxies. The dotted
lines indicate the 98\% error intervals found by bootstrap resampling.
In our bootstrap analysis, we randomly selected $N$ data points, with
substitution, from the original sample of $N$ axis ratios. We then
created a new estimate $\hat{f}$ from the bootstrapped data. After
doing 1000 bootstrap estimates, we then determined the 98\% error
intervals: in Figure~\ref{fig:apparent}, 1\% of the bootstrap
estimates lie above the upper dotted line, and 1\% of the bootstrap
estimates lie below the lower dotted line.

Note from the upper left panel of Figure~\ref{fig:apparent} that
the bright `de' galaxies have a distribution that peaks at
$q_{\rm am} \approx 0.84$, similar to the result found for
relatively bright elliptical galaxies \citep{bg80,fv91,fi91,lm92}.
The faint `de' galaxies have a flatter modal shape ($q_{\rm am}
\approx 0.70$) and a broader distribution of shapes. Both the
bright and faint `de' galaxies have a relative scarcity of
galaxies with $q_{\rm am} = 1$; this is the usual sign that
the galaxies in the population cannot be purely oblate.
For `de' galaxies, the difference between the adaptive moments
shapes and the isophotal shapes is intriguing. Note that
$\hat{f} (q_{25})$ for the `de' galaxies, shown in the
upper right panel of Figure~\ref{fig:apparent}, peaks at
$q_{25} \approx 0.76$ for the bright galaxies.
For the fainter `de' galaxies, the distribution of isophotal axis
ratios, though it peaks at $q_{25} \approx 0.74$, is very broad.
An axis ratio distribution $\hat{f}$ that is nearly constant
over a broad range of $q$ is the signature of a population
of flattened disks. The distribution $\hat{f} (q_{25})$
drops at $q \la 0.2$, indicating that the disks are
not infinitesimally thin, and at $q \ga 0.9$, indicating
that the disks are not perfectly circular.

The distribution of shapes for `de/ex' galaxies are seen
in the second row of Figure~\ref{fig:apparent}. As with
all profile types, the bright galaxies are rounder on
average than the fainter galaxies. Both bright and faint
`de/ex' galaxies have a scarcity of nearly circular galaxies.
Particularly noteworthy is the distinct difference between
$\hat{f} (q_{\rm am})$ for the bright `de' galaxies, very
strongly peaked at $q_{\rm am} \approx 0.84$, and the
equivalent distribution $\hat{f} (q_{\rm am})$ for the
bright `de/ex' galaxies, which is much broader and peaks
at the flatter shape of $q_{\rm am} \approx 0.67$.

The `ex/de' and `ex' galaxies in our sample have, by
definition, $\texttt{fracDeV} < 0.5$ and hence
S\'ersic index $n \la 2$. We expect these galaxies
to be predominantly disk-dominated spiral galaxies.
Determining the true distribution of intrinsic
axis ratios for spiral galaxies is complicated by
the presence of dust. A magnitude-limited sample of
dusty spiral galaxies will show a deficit of galaxies
with high inclination and low apparent axis ratio.
This deficit is due to internal extinction by dust;
in the $B$ band, for instance, a spiral galaxy of
type Sc will be up to 1.5 mag fainter when seen edge-on
than when seen face-on \citep{hv92}. In the $r$ band,
the inclination-dependent dimming is not as great,
but will still affect the observed distribution of
axis ratios in a magnitude-limited sample such as the
SDSS spectroscopic galaxy sample \citep{ry04}. A dusty
disk galaxy which would qualify as a bright `ex' galaxy
if seen face-on might only qualify as a faint `ex' galaxy
when seen edge-on. Similarly, a less luminous galaxy, which
would qualify only as a faint `ex' galaxy when seen face-on,
might drop out of our sample entirely when seen edge-on.
The main effect of dust on the distribution of $q$ for a
magnitude-limited survey of spiral galaxies is to reduce
the number of galaxies with $q \la 0.5$; the distribution
for $q \ga 0.5$ is not strongly affected \citep{hv92}.

With the caveat that we are undercounting edge-on, low-$q$
galaxies, the observed axis ratios for `ex/de' and `ex'
galaxies are shown in the bottom two rows of Figure~\ref{fig:apparent}.
The shapes of `ex/de' galaxies, as measured by either
$q_{\rm am}$ or $q_{25}$, spread over a wide range
of axis ratios, with bright `ex/de' galaxies being slightly
rounder, on average, than the fainter galaxies. The
shapes of `ex' galaxies are very similar to the shapes
of `ex/de' galaxies. For the `ex/de' and `ex' galaxies,
the dip in $\hat{f}$ at $q \ga 0.9$ is a consequence
of the fact that the disks of these galaxies are not
perfectly circular.

\section{INTRINSIC SHAPES}
\label{sec-int}

Having a smooth kernel estimate $\hat{f}$ for the distribution of
apparent axis ratios $q$ permits us to estimate the distribution of 
intrinsic axis ratios $\gamma$, given the assumption that all
galaxies have the same value of the triaxiality parameter $T$.
We don't necessarily expect all the galaxies of a given profile
type to have the same triaxiality; cosmological simulations, as
well as simulations of merging galaxies, generally give a range
of $T$ (see \citet{se04} and references therein) for bright galaxies.
However, by assuming uniform triaxiality and seeing which values
of $T$ yield physically plausible results, we can get a feel
for which values of $T$ are data-friendly, and which are
incompatible with the data.
Let $\hat{N}_T (\gamma)$ be the estimated distribution of
intrinsic axis ratios, given an assumed value for $T$. The
relation between $\hat{f} (q)$ and $\hat{N}_T (\gamma)$
is \citep{bi85}:
\begin{equation}
\hat{f} (q) = \int_0^q P_T ( q | \gamma ) N_T (\gamma) d\gamma \ ,
\label{eq:volterra}
\end{equation}
where $P_T ( q | \gamma ) dq$ is the probability that a galaxy
with intrinsic short-to-long axis ratio $\gamma$ and triaxiality
$T$ has an apparent axis ratio in the range $q \to q + dq$, when
viewed from a random orientation.

Equation~(\ref{eq:volterra}) represents a Volterra
equation of the first kind which can be solved by standard
numerical techniques. We first compute $\hat{f}$ on 200 points equally
spaced in $q$, using the kernel technique previously
described. We then compute $P_T$ on an $200 \times 200$ grid
in $(q,\gamma)$ space. For each value of $\gamma$, we compute
an adaptive kernel estimate for $P_T$ by randomly choosing a large
number of viewing angles ($\sim 10^5$), then computing
what the apparent axis ratio $q$ would be for each viewing angle,
given the assumed values of $\gamma$ and $T$ \citep{bi85}. The kernel
estimate of the distribution of $q$ is then our estimate
of $P_T ( q | \gamma )$. Equation~(\ref{eq:volterra}), in its
discretized form, can be written
\begin{equation}
\hat{f}_k = \sum_{j = 1}^k P_{T,kj} \hat{N}_{T,j} \ .
\label{eq:matrix}
\end{equation}
Since $P_{T,kj} = 0$ for $j > k$ (the apparent axis ratio of
a galaxy can't be less than its intrinsic axis ratio),
the matrix $\mathbf{P}_T$ is lower triangular,
and equation~(\ref{eq:matrix}) can be inverted simply by
forward substitution \citep{pr92}.

We performed the inversion to find $\hat{N}_T$ using
six different values of the trixiality:
$T = 0$ (oblate), $T = 0.2$, $T = 0.4$, $T = 0.6$,
$T = 0.8$, and $T = 1$ (prolate). When our ultimate goal
is finding $\hat{N}_T$, determining the optimal kernel
width $h$ for $\hat{f}$ is a vexed question. It is
generally true that the best value of
$h$ for estimating $\hat{N}_T$ is greater than the
best value of $h$ for purpose of estimating $\hat{f}$ \citep{tm95,tm96};
$N_T$ is a deconvolution of $f$, and the deconvolution
process increases the noise present. 
\citet{tm95} proposed, for the purposes of estimating
$N_T$, using the value of $h$ that minimized
\begin{equation}
{\rm BMISE} (h) = \int_0^1 [ \hat{N}_T (\gamma) - \hat{N}_t^* (\gamma) ]^2 d\gamma \ ,
\end{equation}
where $\hat{N}_T^*$ is an estimate of $N_T$ from a bootstrap
resampling of the original data. Unfortunately, we found, as
\citet{tm95} did, that ${\rm BMISE} (h)$ returns
badly oversmoothed estimates of $N_T$. Thus, we fell
back on using our subjective impressions of the smoothness
of $\hat{N}_T$, and ended by taking the value of $h$
given by equation~(\ref{eq:hwidth}) and multiplying by a
factor of 1.5.

To perform the inversion and find the estimated distribution
of $\gamma$, we assumed that all galaxies have the
same triaxiality $T$. If this assumption is incorrect, then
the inversion of equation~(\ref{eq:matrix}) may result in
$\hat{N}_T < 0$ for some range of $\gamma$, which is
physically impossible. To exclude our hypothesis (that
all galaxies have triaxiality $T$) at some fixed statistical
confidence level, we can repeat the inversion for a large number
of bootstrap resamplings of the original data set. In practice,
we did 1000 resamplings, and used them to place 98\% confidence
intervals on $\hat{N}_T$. If the confidence interval falls below
zero for some valus of $\gamma$, then the hypothesized value of
$T$ can be rejected at the 99\% (one-sided) confidence level.

\begin{figure}
\plotone{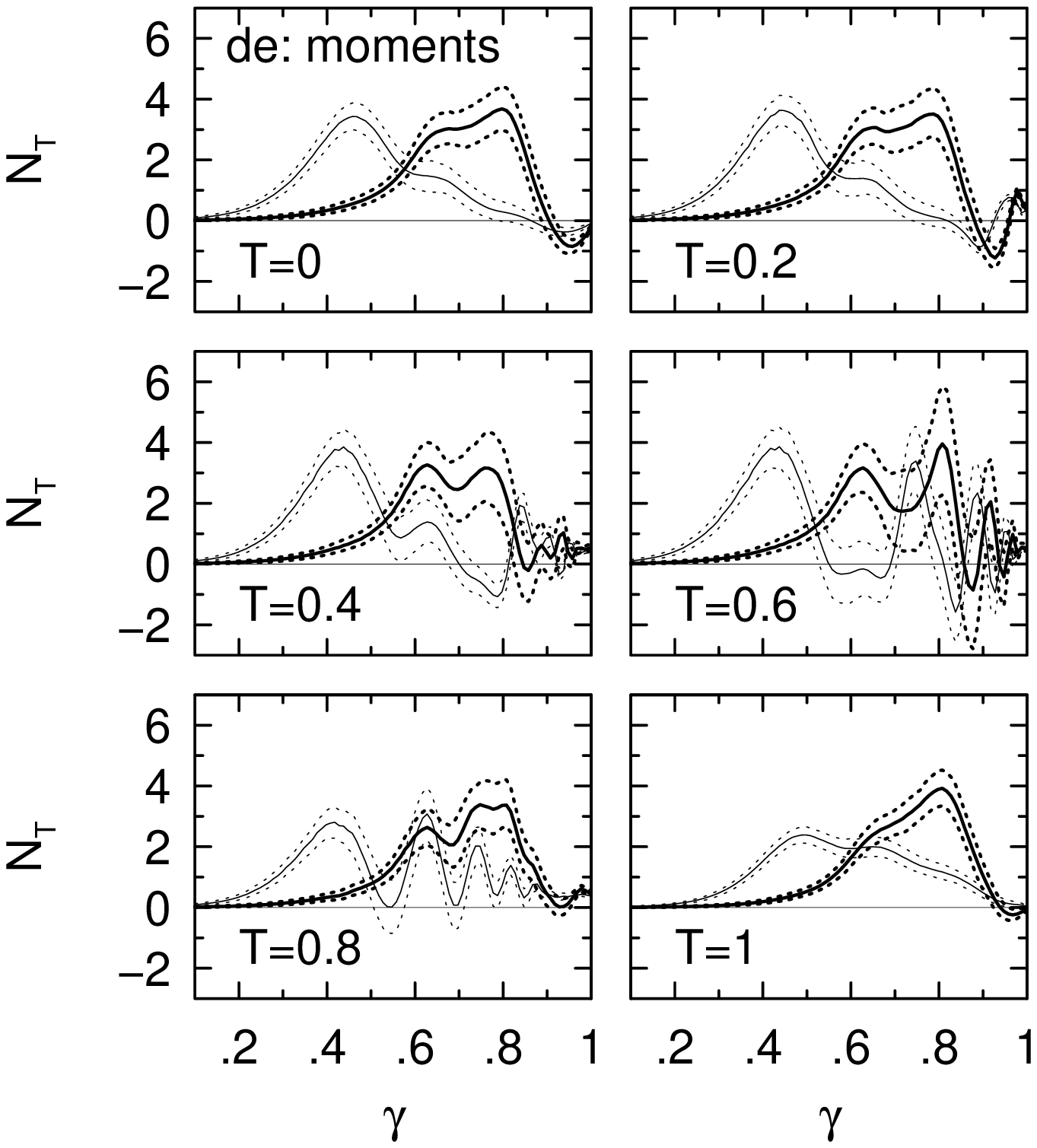}
\caption{
Distribution of intrinsic axis ratios for galaxies
of profile type `de' (\texttt{fracDeV} $> 0.9$), using
the adaptive moments estimate of $q$. The heavy line is the distribution
for galaxies with $M_r \leq -21.8$; the
light line is for galaxies with $M_r > -21.8$.
The dotted lines indicate the 98\% confidence intervals,
estimated by bootstrap resampling. The assumed value of
the triaxiality $T$ is given in each panel.
}
\label{fig:de_adapt}
\end{figure}

Figure~\ref{fig:de_adapt} shows the distribution of
intrinsic shapes for `de' galaxies ($\texttt{fracDeV} > 0.9$),
when the adaptive moments axis ratio $q_{\rm am}$ is used as
the estimator of the apparent shape. 
The heavy line is the distribution for the bright `de' galaxies ($M_r \leq
-21.8$) and the light line is the distribution for the fainter `de' galaxies.
The assumed value of $T$ is shown in each panel, ranging from purely
oblate shapes, with $T=0$, in the upper left corner to purely prolate
shapes, with $T = 1$, in the lower right corner.

An eyecatching feature of Figure~\ref{fig:de_adapt} is the oscillatory
nature of $\hat{N}_T$ in the triaxial cases. The multiple peaks in
$\hat{N}_T$ at high $\gamma$
result from the shape of the conditional probability
function $P_T (q | \gamma )$ for highly triaxial galaxies. When
galaxies are axisymmetric ($T = 0$ or $1$), the conditional probability
peaks at $q = \gamma$, the minimum possible apparent axis ratio.
For triaxial galaxies, however, the conditional probability
has one or more local maxima at $q > \gamma$. To see why this
can result in oscillatory solutions consider $\hat{N}_T$ for
faint `de' galaxies when $T = 0.8$; this is
shown by the light line in the lower left panel
of Figure~\ref{fig:de_adapt}. The distribution $\hat{N}_T$ has a local
maximum at $\gamma = 0.42$. For $T = 0.8$, the conditional probability
$P_T (q | \gamma = 0.42)$ peaks at $q = 0.57$. Thus, the
large number of $\gamma \approx 0.42$ galaxies result in
many galaxies with an apparent axis ratio $q \approx 0.57$; 
so many, in fact, that
$\hat{N}_T$ must be made very small in the range $0.42 < \gamma
< 0.57$ in order to avoid overproducing $q \approx 0.57$ galaxies.
In fact, as seen in Figure~\ref{fig:de_adapt}, $\hat{N}_T$
has a local minimum at $\gamma = 0.54$, at which $\hat{N}_T$
falls slightly below zero. Since $P_T (q | \gamma = 0.54)$
peaks at $q = 0.65$, this produces a deficit of $q \approx 0.65$
galaxies, which must be compensated for by making $\hat{N}_T$
very large in the range $0.54 < q < 0.65$. In fact, $\hat{N}_T$
has another local maximum at $q = 0.62$. And so the oscillations
continue, with decreasing wavelength, until $\gamma = 1$ is reached.
\citet{tm95} found that $\hat{N} (\gamma)$ for bright elliptical
galaxies was significantly bimodal if the galaxies were assumed to be
highly triaxial; this bimodality also had its origin in the shape
of the conditional probability function for triaxial galaxies.

For both bright and faint `de' galaxies, using $q_{\rm am}$
as the apparent axis ratio, oblate fits are statistically
unacceptable, producing a negative number of galaxies with
$\gamma \ga 0.9$ (see the upper left panel of Figure~\ref{fig:de_adapt}).
For bright `de' galaxies, statistically acceptable fits
are found for $T = 0.4$, $0.6$, and $0.8$; that is, their
98\% confidence intervals never fall entirely below zero.
The mean intrinsic axis ratio for the acceptable fits ranges
from $\langle \gamma \rangle = 0.66$ for $T = 0.4$ to
$\langle \gamma \rangle = 0.69$ for $T = 0.8$. The
permissible fits for the bright `de' galaxies are
consistent with the deduced triaxial shape of the nearby bright
($M_r \sim -22.2$) elliptical galaxy NGC 4365, for
which a combination of photometric and kinematic
data yields $T \sim 0.45$ and $\gamma \sim 0.6$
\citep{se04}. Although
the bright `de' galaxies are best fit by highly triaxial
shapes, the faint `de' galaxies are best fit by nearly prolate
shapes, with $T = 0.8$ and $T = 1$ both giving statistically
acceptable fits. (The highly oscillatory solution for
$T = 0.8$ may be physically dubious -- why should galaxy
shapes be quantized? -- but it is statistically acceptable.)
The mean intrinsic shape of faint `de' galaxies is
$\langle \gamma \rangle = 0.53$ if $T = 0.8$ and $\langle
\gamma \rangle = 0.58$ if $T = 1$.

\begin{figure}
\plotone{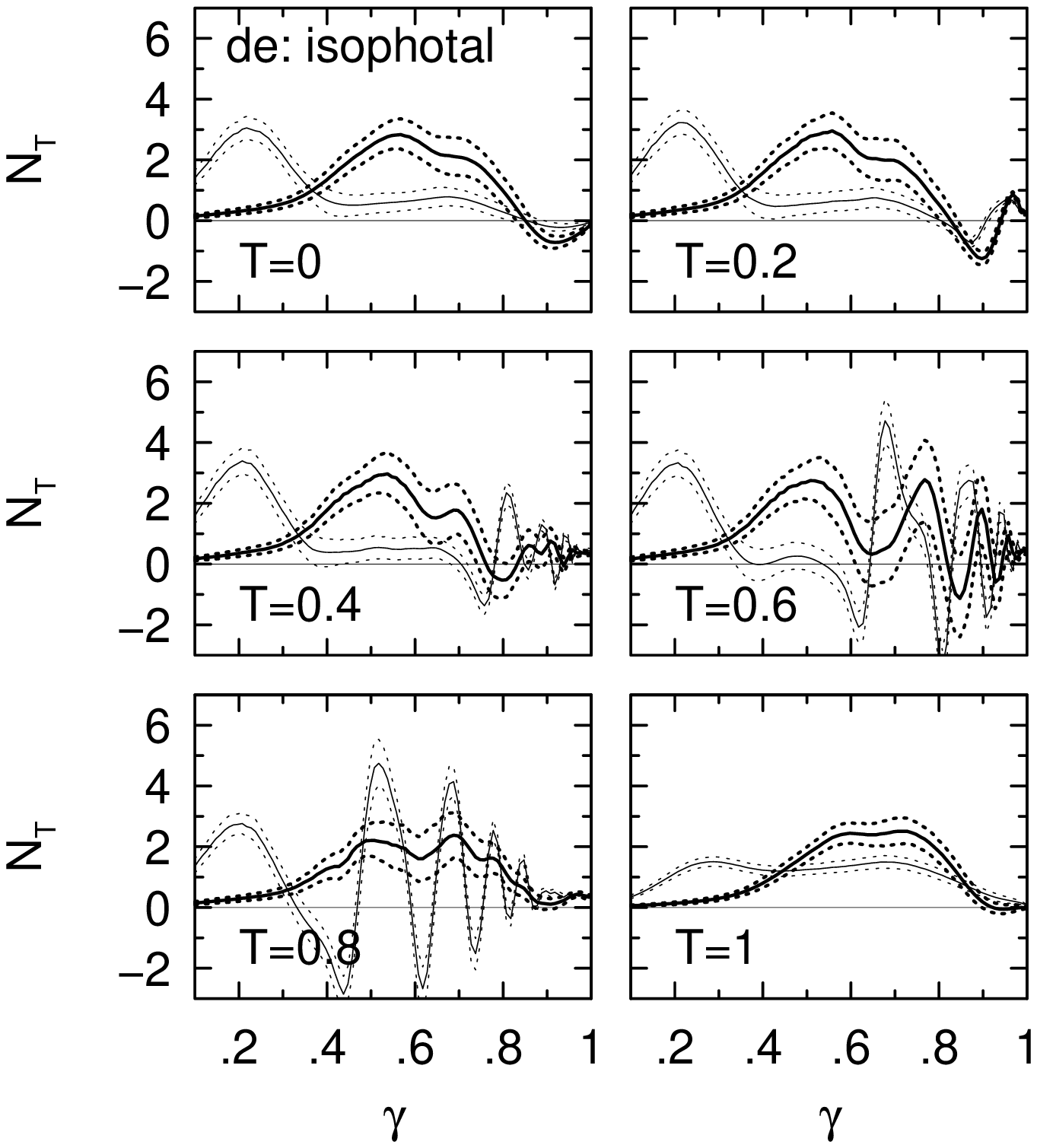}
\caption{Same as Figure~\ref{fig:de_adapt}, but using the apparent
axis ratio of the 25 mag/arcsec$^2$ isophote as the shape estimate.
}
\label{fig:de_iso}
\end{figure}

Figure~\ref{fig:de_iso} shows the deduced distribution of
intrinsic shapes for `de' galaxies when the isophotal
axis ratio $q_{25}$ is used, rather than the adaptive
moments axis ratio. The bright `de' galaxies have a statistically
acceptable fit when $T = 0.4$, $T = 0.6$, $T = 0.8$, and
$T = 1$. The mean intrinsic shape ranges from $\langle \gamma
\rangle = 0.51$ when $T = 0.4$ to $\langle \gamma \rangle = 0.62$
when $T = 1$. The fainter `de' galaxies are acceptably
fit, assuming constant triaxiality, only when
$T = 1$, which results in $\langle \gamma \rangle = 0.51$. The
$T = 0$ case, which can be rejected at the 99\% confidence
level but not at the 99.9\% level, would produce $\langle \gamma
\rangle = 0.28$; this axis ratio is flatter than that
of an ice hockey puck. (Although the $T = 0$ and $T = 0.2$ fits
are statistically unacceptable at the 99\% confidence level,
the data are consistent with a population of nearly oblate
shapes if we relax the assumption of uniform $T$. For
instance, the faint `de' galaxies can be fit at the
99\% confidence level, with a Gaussian distribution of $T$
peaking at $T = 0$ and with $\sigma_T = 0.2$.)

\begin{figure}
\plotone{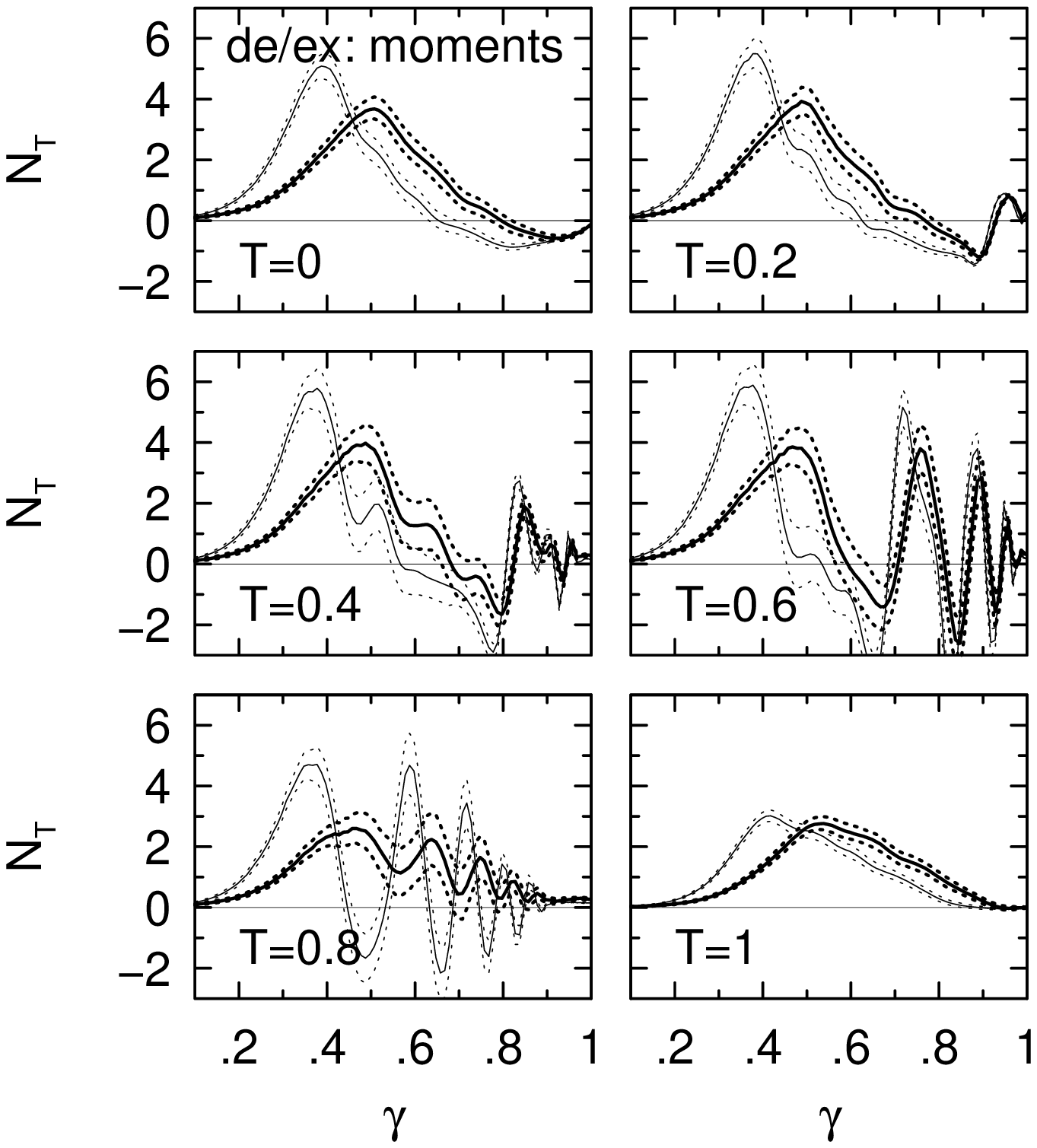}
\caption{Distribution of intrinsic axis ratios for
galaxies of profile type `de/ex' ($0.5 <$ \texttt{fracDeV} $\leq 0.9$),
using the adaptive moments estimate of $q$. The heavy line is the
distribution for galaxies with $M_r \leq -21.35$;
the light line is for galaxies with $M_r > -21.35$.
The dotted lines indicate the 98\% confidence intervals,
estimated by bootstrap resampling.
The assumed value of the triaxiality $T$ is given in each panel.
}
\label{fig:deex_adapt}
\end{figure}

The shapes of `de/ex' galaxies can be analyzed in the same way
as the shapes of `de' galaxies. For instance, the distributions of
intrinsic shapes for `de/ex' galaxies, using $q_{\rm am}$ as the
apparent axis ratio, are shown in Figure~\ref{fig:deex_adapt}.
For the brighter galaxies with `de/ex' profiles, acceptable
fits are found when $T = 0.8$, yielding $\langle \gamma \rangle
= 0.51$, and when $T = 1$, yielding $\langle \gamma \rangle = 0.57$.
For the fainter `de/ex' galaxies, only the $T = 1$ fit is
statistically acceptable, yielding a mean intrinsic axis
ratio of $\langle \gamma \rangle = 0.48$ for the prolate galaxies.
Notice that bright `de/ex' galaxies, faint `de/ex' galaxies, and
faint `de' galaxies all are consistent with having shapes, at
least in their inner regions, which are prolate, or nearly prolate,
with a typical axis ratio of $\langle \gamma \rangle \sim 1/2$.
Bright `de' galaxies are distinctly different, with shapes that
are consistent with their being highly triaxial, with a typical
axis ratio of $\langle \gamma \rangle \sim 2/3$.

\begin{figure}
\plotone{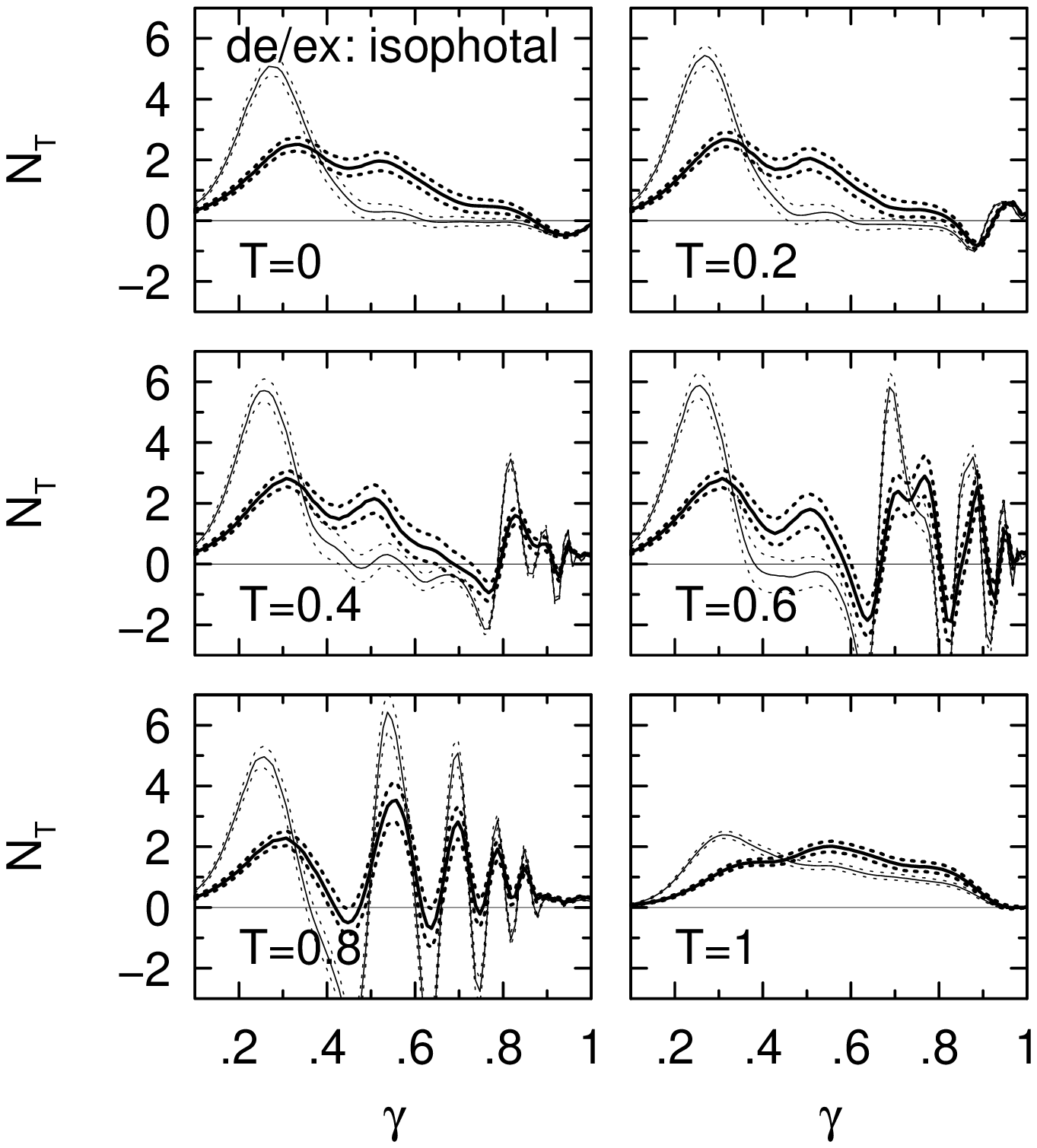}
\caption{Same as Figure~\ref{fig:deex_adapt}, but using the apparent
axis ratio of the 25 mag/arcsec$^2$ isophote as the shape estimate.
}
\label{fig:deex_iso}
\end{figure}

The intrinsic shape distributions for `de/ex' galaxies, using
$q_{25}$ as the apparent shape measure, are shown in
Figure~\ref{fig:deex_iso}. For both the bright and faint
`de/ex' galaxies, the $T = 1$ fit is statistically acceptable.
For the prolate fits, the average shape of the bright `de/ex'
galaxies is $\langle \gamma \rangle = 0.55$, and the average
shape of the faint `de/ex' galaxies is $\langle \gamma \rangle
= 0.47$. Thus, given the prolate hypothesis, `de/ex' galaxies
are only slightly flatter in their outer regions, whose shape
is measured by $q_{25}$, than in their inner regions, whose
shape is measured by $q_{\rm am}$.

\begin{figure}
\plotone{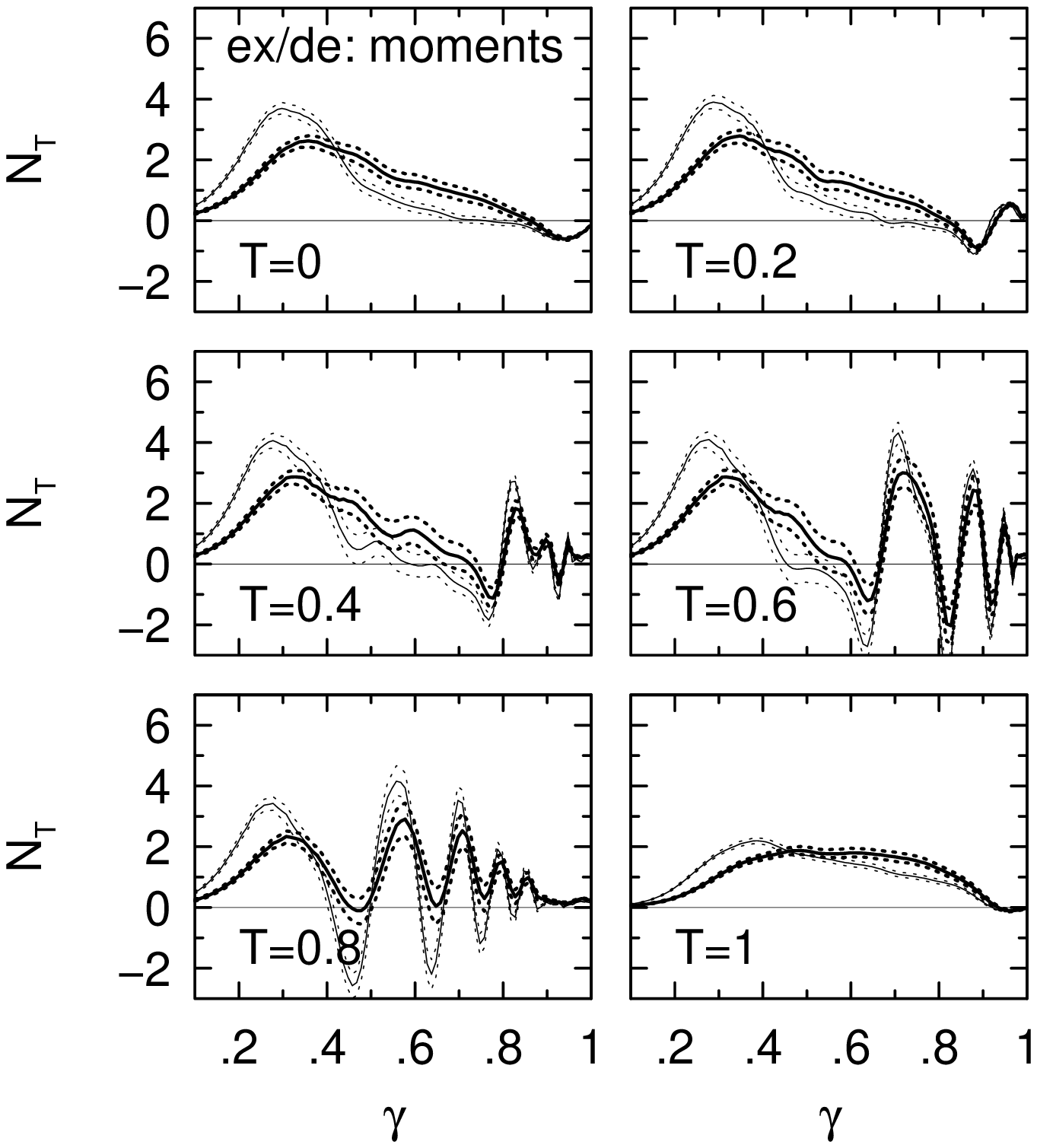}
\caption{Distribution of intrinsic axis ratios for
galaxies of profile type `ex/de' ($0.1 <$ \texttt{fracDeV} $\leq 0.5$),
using the adaptive moments estimate of $q$. The heavy line is the
distribution for galaxies with $M_r \leq -21.0$;
the light line is for galaxies with $M_r > -21.0$.
The dotted lines indicate the 98\% confidence intervals,
estimated by bootstrap resampling.
The assumed value of the triaxiality $T$ is given in each panel.
}
\label{fig:exde_adapt}
\end{figure}

\begin{figure}
\plotone{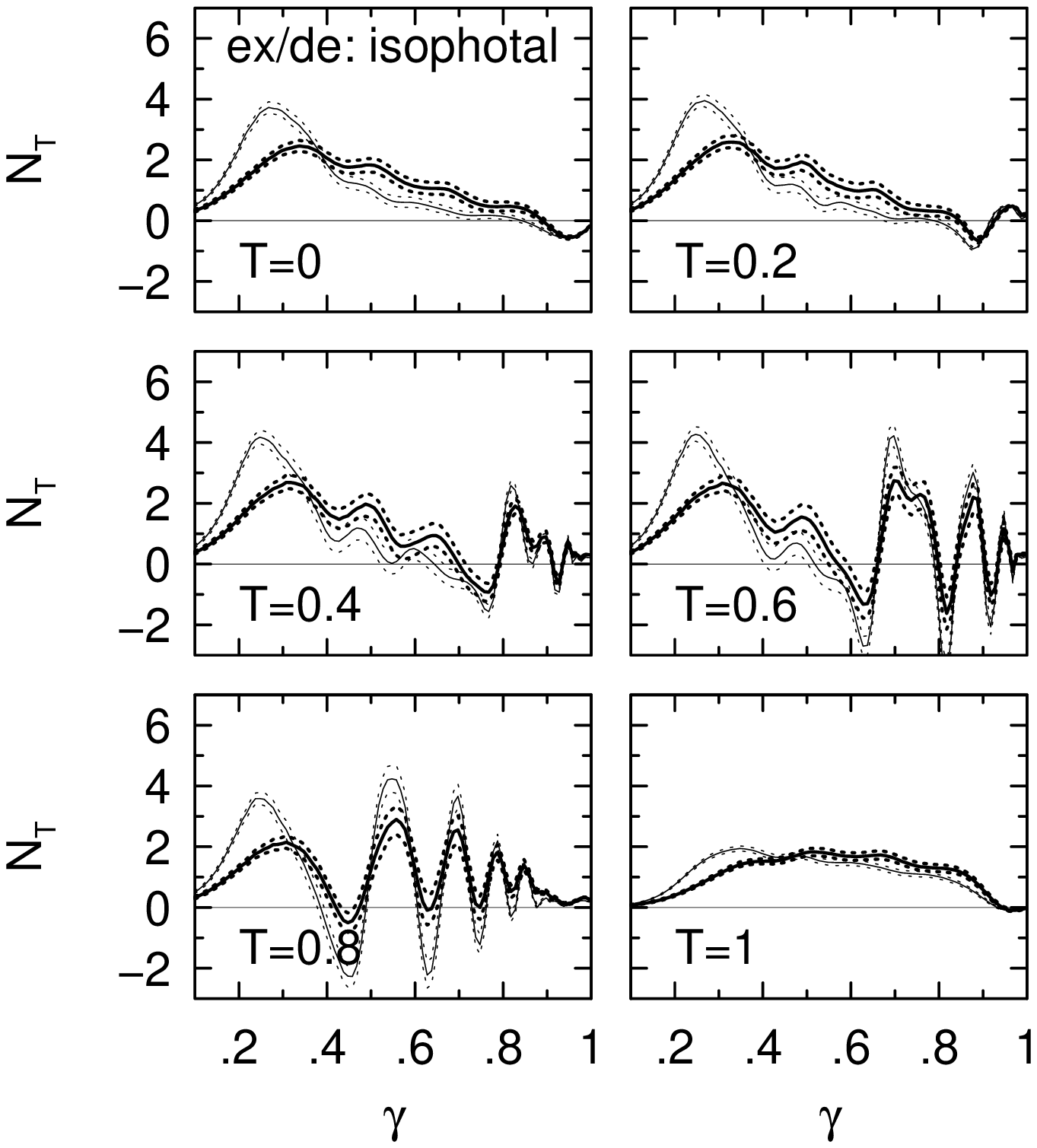}
\caption{Same as Figure~\ref{fig:exde_adapt}, but using the apparent
axis ratio of the 25 mag/arcsec$^2$ isophote as the shape estimate.
}
\label{fig:exde_iso}
\end{figure}

With the repeated caveat that we are undercounting edge-on, low-$q$ galaxies,
the results for the `ex/de' galaxies are presented
in Figures~\ref{fig:exde_adapt} and \ref{fig:exde_iso}.
When $q_{\rm am}$ is used as the measure of apparent shape
(Figure~\ref{fig:exde_adapt}), the bright `ex/de' galaxies
have a statistically acceptable fit when $T = 0.8$, which
results in $\langle \gamma \rangle = 0.49$; the $T = 1$ fit,
which goes slightly negative at $\gamma \ga 0.94$, yields
a mean intrinsic axis ratio of $\langle \gamma \rangle = 0.55$.
None of the six tested triaxialities gave an
acceptable fit, at the 99\% confidence level, to
the faint `ex/de' galaxies. The best (or the `least bad') of
the fits, with $T = 1$, yields $\langle \gamma \rangle = 0.49$.
When $q_{25}$ is used as the measure of apparent shape
(Figure~\ref{fig:exde_iso}), none of the tested values of
$T$ gives a fit acceptable at the 99\% confidence level.
The best fit, for both bright and faint galaxies, is given
by $T = 1$, for which $\langle \gamma \rangle = 0.55$
for the bright `ex/de' galaxies and $\langle \gamma \rangle
= 0.50$ for the faint `ex/de' galaxies.

\begin{figure}
\plotone{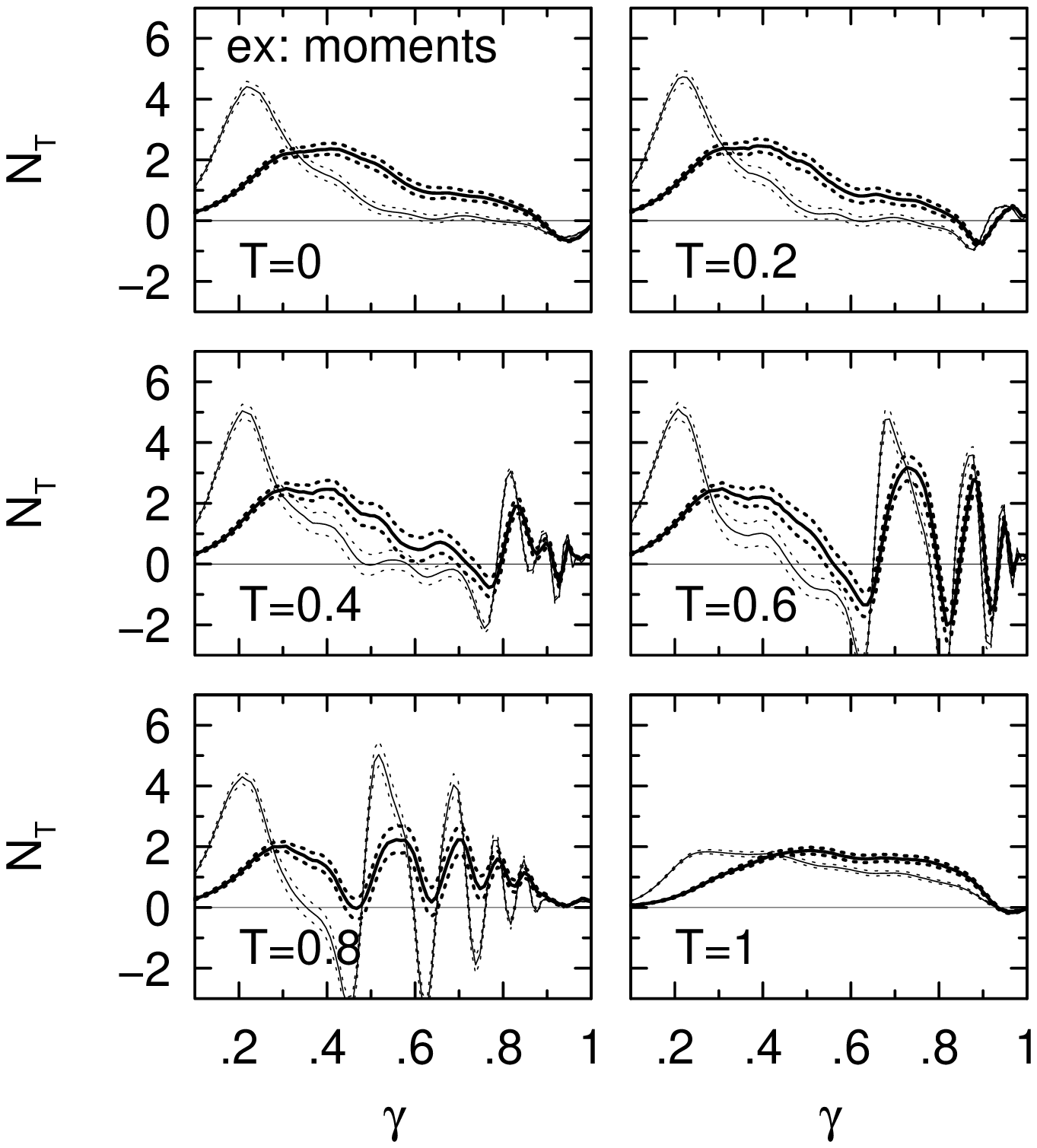}
\caption{Distribution of intrinsic axis ratios for
galaxies of profile type `ex' (\texttt{fracDeV} $\leq 0.1$), using the
adaptive moments estimate of $q$. The heavy line is the distribution
for galaxies with $M_r \leq -20.5$; the light line
is for galaxies with $M_r > -20.5$.
The dotted lines indicate the 98\% confidence intervals,
estimated by bootstrap resampling.
The assumed value of the triaxiality $T$ is given in each panel.
}
\label{fig:ex_adapt}
\end{figure}

\begin{figure}
\plotone{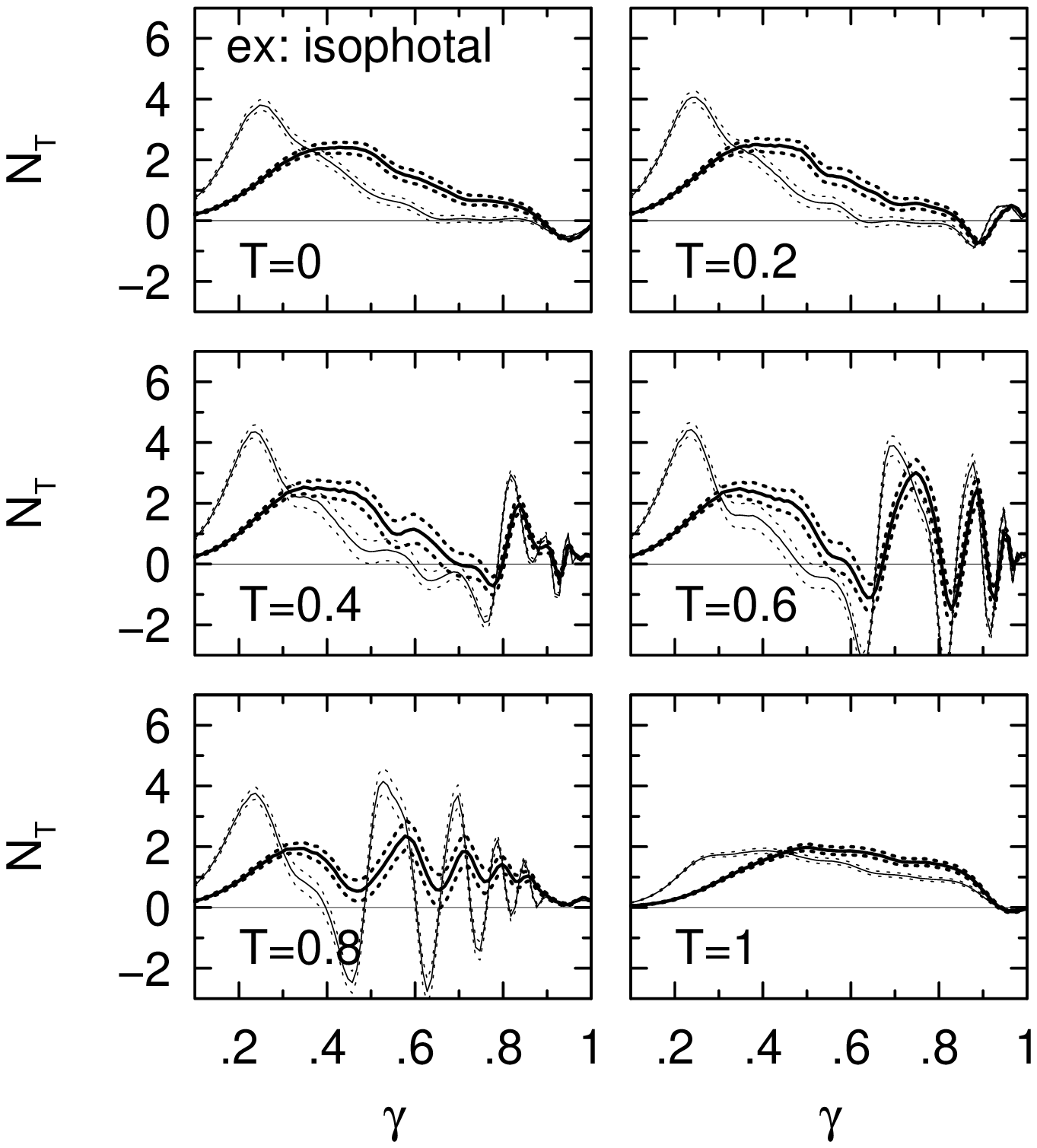}
\caption{Same as Figure~\ref{fig:ex_adapt}, but using the apparent
axis ratio of the 25 mag/arcsec$^2$ isophote as the shape estimate.
}
\label{fig:ex_iso}
\end{figure}

As with the `ex/de' galaxies, the sample of `ex' galaxies is
affected by an undercount of low-$q$ galaxies. With this
caveat, the results for the `ex' galaxies are shown in
Figures~\ref{fig:ex_adapt} and \ref{fig:ex_iso}. In general,
the results are similar to those for `ex/de' galaxies.
When the axis ratios of `ex' galaxies are estimated using
$q_{\rm am}$ (Figure~\ref{fig:ex_adapt}), the bright `ex'
galaxies are acceptably fit when $T = 0.8$, yielding
$\langle \gamma \rangle = 0.49$; the $T = 1$ fit for
the bright `ex' galaxies, which goes slightly negative
at high $\gamma$, yields $\langle \gamma \rangle = 0.56$.
The best fit to the faint `ex' galaxies is provided by
the $T = 1$ inversion, which yields $\langle \gamma \rangle
= 0.46$. When $q_{25}$ is used as the estimate of the
apparent axis ratio (Figure~\ref{fig:ex_iso}), the results
are extremely similar. For bright `ex' galaxies, only $T = 0.8$,
of the tested triaxialities, gave a fit acceptable at the
99\% confidence level, yielding $\langle \gamma \rangle =
0.51$; the not-as-good $T = 1$ fit gave $\langle \gamma \rangle
= 0.57$. The best fit for faint `ex' galaxies had $T = 1$,
$\langle \gamma \rangle = 0.49$.

Although the best fits to the `ex/de' and `ex' galaxies, under
the assumption of constant triaxiality, were prolate or nearly
prolate, we have an a priori knowledge that bright galaxies
with S\'ersic index $n \la 2$ are generally rotationally
supported disks. The SDSS DR3 photometry of `ex/de' and
`ex' galaxies is consistent with their being nearly oblate
objects only if they have a range of $T$, instead of being
forced into a straitjacket of uniform $T$.

\section{DISCUSSION}
\label{sec-dis}

Constraining the flattening $\gamma$ and triaxiality $T$ of galaxies
with different luminosity and profile type provides potentially
useful clues for the study of galaxy formation and evolution.
For example, the dissipationless merger of two equal-mass disk
galaxies embedded within dark halos produces a merger remnant
whose luminous component has an approximate de Vaucouleurs
profile \citep{ba92}. The values of $\gamma$ and $T$ of the remnant
depend on the initial orbital parameters of the merging galaxies,
but the remnant is typically quite flat and prolate. For a set of
eight simulations with different initial orbits, \citet{ba92}
found $\langle \gamma \rangle \sim 0.5$ and $\langle T \rangle
\sim 0.7$ for the most tightly bound quartile of the luminous
particles (the inner region of the galaxy). This is significantly
flatter and more prolate than the shape we deduced from the
distribution of $q_{\rm am}$ for bright `de' galaxies. The
oblateness of the merger remnant is increased, however, if
the initial merging galaxies are unequal in mass \citep{ba98,nb03}.

Including gaseous dissipation into a simulated merger can
strongly affect $\gamma$ and $T$. As the gas dissipates and
falls toward the center of the merger remnant, the central
mass concentration will scatter stars on low angular momentum
box orbits, thus acting to increase $\gamma$ and decrease $T$.
For instance, a simulated remnant that has
$\gamma \approx 0.55$ and $T \approx 0.8$
in the absence of dissipation will have
$\gamma \approx 0.65$ and $T \approx 0.5$ if gas
dissipation is added \citep{ba98}. The deduced shapes
of bright `de' galaxies are consistent with their being
merger remnants, as long as the mergers were dissipational.

In the standard hierarchical clustering model for structure
formation, however, we do not expect most elliptical galaxies
to have formed by the relatively recent merger of a pair
of spiral galaxies. More typically, an elliptical galaxy
will have formed by the successive merger of a number of
smaller galaxies or subgalactic ``clumps''. Simulations
of multiple mergers in small groups reveal that the final
merger remnant generally has a surface brightness profile
similar to a de Vaucouleurs law, but that the shape of
the merger remnant depends on the assumed initial conditions
\citep{wh96,ga97,zw02}. For instance, \citet{zw02} simulated
dissipationless mergers of dense clumps embedded in a smooth
dark halo. The profile of the merger remnant, in all cases,
was well fit by a de Vaucouleurs law. The triaxiality of
the remnant depended primarily on whether the system was
initially in a state of virial equilibrium or in a state
of non-virial collapse. The virial case produced more oblate
remnants ($\langle T \rangle \sim 0.3$), while the collapse
case produced more prolate remnants ($\langle T \rangle \sim 0.7$).
The virial case also produced less flattened remnants:
$\langle \gamma \rangle \sim 0.8$ for the merger remnants
of initially virialized systems versus $\langle \gamma \rangle
\sim 0.6$ for initially collapsing systems. The value of
$\gamma$ also depended on the initial clump-to-halo mass
ratio; if the clumps contributed only a small fraction of
the total mass, they produced a flatter remnant than if they
contributed all the mass, with no dark halo.
The best fitting parameters for the bright `de' galaxies,
$\mu_\gamma = 0.66$ and $T_0 = 0.43$, suggest that they are too
flattened and too triaxial to have been formed primarily by mergers
within a virialized group, which produce $\langle \gamma \rangle
\sim 0.8$ and $\langle T \rangle \sim 0.3$. (Remember, the
effects of dissipation will only act to increase $\gamma$
and decrease $T$.)

Another influence that is capable of increasing $\gamma$ and decreasing
$T$ is the presence of a central supermassive black hole. A central
mass concentration affects the structure of the surrounding galaxy
by disrupting box orbits \citep{gb85,nm85}. The resulting chaotic
orbits lead to an altered shape for the galaxy; the trend is
toward shapes that are nearly spherical in the inner regions
and nearly oblate in the outer regions \citep{mq98}. In the
inner regions, the orbits can support a shape that is
nearly oblate, $T \sim 0.25$, or strongly triaxial,
$T \sim 0.5$, but not nearly prolate, $T \sim 0.75$ \citep{pm04}.
The time scale for the shape evolution depends on the mass of
the central black hole relative to the total stellar mass of the galaxy.
When the central black hole has a mass equal to 1\% of the
mass in stars, its shape at the half-mass radius
evolves in $\sim 40$ times the half-mass orbital period;
when the mass is only 0.3\% of
the stellar mass, the evolution time scale increases to
$\sim 200$ times the orbital period.
The observed relation between central black hole mass
and bulge mass \citep{mf01, mh03, hr04} predicts that an
elliptical galaxy will typically have a central black
hole with a mass $\la 0.2\%$ of the total mass in stars.
For such relatively low black hole masses, the
shape evolution time scale at the half-mass radius
and beyond will exceed the age of the galaxy.
Thus, a central black hole will not generally affect
the isophotal axis ratio $q_{25}$.

The isophotal shape, $q_{25}$, measures the apparent shape of a
galaxy at a few times the effective radius. Converted to physical
units, the average semimajor axis $A_{25}$ of the isophotal ellipse
ranges from $\langle A_{25} \rangle = 2.4 \kpc$ at $M_r = -16$
to $\langle A_{25} \rangle = 45 \kpc$ at $M_r = -23$.
At relatively high luminosities ($M_r \la -20.5$),
the mean value of $q_{25}$ increases with luminosity;
bright galaxies are rounder than fainter galaxies at
their outer regions. We have also uncovered the puzzling result
that at a given luminosity, the mean value of $q_{25}$ is larger
for `ex' galaxies than for `de' galaxies.
Why should `ex' galaxies,
which are flattened disks in their central regions, be rounder
than `de' galaxies in their outskirts? At such great distances
from the galaxy's center, we expect that tidal distortions from
neighboring galaxies will significantly affect the isophotal shape.
Galaxies in dense regions, such as the cores of clusters, are subject
to more interactions with neighboring galaxies, and may well differ
systematically in shape from field galaxies, which are less
frequently harassed. Since elliptical galaxies are found
preferentially in dense environments, this conjectured environment --
shape relation would then translate into a relation between
profile type and shape. A future paper, studying the relation
between apparent axis ratio and environment for SDSS DR3 galaxies
(Kuehn \& Ryden, 2005, in preparation), will
test this conjecture.

\acknowledgments

Richard Pogge, Fred Kuehn, and the
anonymous referee provided useful assistance and comments.

Funding for the creation and distribution of the SDSS Archive has been
provided by the Alfred P. Sloan Foundation, the Participating Institutions,
the National Aeronautics and Space Administration, the National Science
Foundation, the U.S. Department of Energy, the Japanese Monbukagakusho,
and the Max Planck Society. The SDSS website is \url{http://www.sdss.org/}.
The SDSS is managed by the Astrophysical Research Consortium (ARC) for
the Participating Institutions. The Participating Institutions are
The University of Chicago, Fermilab, the Institute for Advanced Study,
the Japan Participation Group, The Johns Hopkins University, Los Alamos
National Laboratory, the Max-Planck-Institute for Astronomy (MPIA), the
Max-Planck-Institute for Astrophysics (MPA), New Mexico State University,
University of Pittsburgh, Princeton University, the United States Naval
Observatory, and the University of Washington.

\clearpage

\begin{deluxetable}{lcccr}
\tablewidth{0pt}
\tablecaption{Dividing Line Between Bright and Faint Galaxies\label{tab:bridim}}
\tablehead{
\colhead{Profile Type} & \colhead{$M_0$} & \colhead{$N(M_r\leq M_0)$} &
\colhead{$N(M_r > M_0)$} & \colhead{$\log P_{\rm KS}$} }
\startdata
de    & -21.84 & \phn\phn  926 & \phn 1311 &  -52.9 \\
de/ex & -21.35 & \phn 5235 & \phn 8545 &  -109.2 \\
ex/de & -21.00 & 11340 & 25305 &  -118.6 \\
ex    & -20.48 & 17101 & 27188 & -301.9 \\
\enddata
\end{deluxetable}

\end{document}